\documentclass[twocolumn,aps,prl,showpacs,groupedaddress,superscriptaddress,%
nofootinbib,floatfix]{revtex4}%
\usepackage{graphicx}
\usepackage{amsmath}
\usepackage{bm}
\begin{document}
\title{Optimal spin-quantization axes for the polarization\\
	of dileptons with large transverse momentum 
}
\author{Eric~Braaten}
\affiliation{Physics Department, Ohio State University, 
Columbus, Ohio 43210, USA}
\author{Daekyoung~Kang}
\affiliation{Physics Department, Ohio State University, 
Columbus, Ohio 43210, USA}
\author{Jungil~Lee}
\affiliation{Department of Physics, Korea University, Seoul 136-701, Korea}
\author{Chaehyun~Yu}
\affiliation{Department of Physics, Korea University, Seoul 136-701, Korea}

\begin{abstract}
The leading-order parton processes that produce a dilepton 
with large transverse momentum predict that 
the transverse polarization should increase with the transverse momentum 
for almost any choice of the quantization axis 
for the spin of the virtual photon.  The rate of approach to complete 
transverse polarization depends on the choice of spin quantization axis.
We propose axes that optimize that rate of approach.
They are determined by the momentum of the dilepton and the direction 
of the jet that provides most of the balancing transverse momentum.
\end{abstract}
\pacs{13.88.+e,13.85.Qk,13.90.+i}

\maketitle

The spins of particles produced in high energy collisions 
carry important information about the fundamental interactions 
between elementary particles, but it is difficult to access 
that information.  Since the spin cannot be measured directly,
information about the spin of a particle must be inferred 
from the angular distribution of its decay products.
The accessibility of information about the spin
is affected by the choice of a spin quantization axis (SQA),
because some angular variables are more easily measured than others.
Hadron collisions provide an additional complication, because the
fundamental interactions involve collisions of partons with 
varying longitudinal momenta.  Integration over these longitudinal 
momenta tends to dilute the information carried by the spins of the 
final-state particles.  This raises an important question:
which SQA will maximize that information?

\vskip -1ex
The simplest process for which this question can be addressed 
is the production of a {\it dilepton}, 
a lepton and antilepton with opposite charges.
There are various sources of dileptons in high energy collisions, 
but we focus on production through a virtual photon, 
which can be regarded as a spin-1 particle with a variable mass 
equal to the invariant mass of the lepton pair.
There are some simple dilepton production mechanisms
that predict that the virtual photon should
be transversely polarized.  For the 
{\it Drell-Yan mechanism}, the annihilation of a quark and an antiquark 
into a virtual photon, the vector coupling to the quark and 
antiquark ensures that the virtual photon is transversely 
polarized \cite{Drell:1970wh}.  The leading-order parton processes 
for a dilepton with large transverse momentum are
dominated asymptotically by {\it photon fragmentation},
the production of a transversely-polarized real photon 
followed by its decay into a collinear lepton pair \cite{Braaten:2001sz}. 
For these examples, the leading-order prediction of perturbative QCD 
is that the virtual photon is transversely polarized.  
Longitudinal virtual photons are 
produced by higher-order processes at a rate that depends on the
SQA.  A reasonable criterion for an optimal 
SQA is that it minimizes the cross section 
for producing a longitudinal virtual photon.

For the Drell-Yan mechanism,
an optimal SQA was identified long ago:
it is the {\it Collins-Soper axis} \cite{Collins:1977iv}.
In this paper, we derive optimal SQA's 
for the parton processes that create a dilepton 
with large transverse momentum.  
We apply them to dilepton production at Fermilab's 
Tevatron and at CERN's Large Hadron Collider (LHC).

We consider the production of a dilepton whose invariant mass $Q$
is well below the $Z^0$ resonance. 
The orientation of the lepton momentum 
in the dilepton rest frame is given by a 
polar angle $\theta$ with respect to a chosen
axis and an azimuthal angle $\phi$ with respect to a chosen plane
containing that axis.  If $\phi$ 
is integrated over, the differential cross section 
can be expressed as
\begin{equation}
\frac{d \sigma}{d x} = \frac{\alpha}{4 \pi} 
\int \!\! \frac{dQ^2}{Q^2}
\bigg[  
 \sigma_T \, \frac{1 + x^2}{2}
    + \sigma_L \, (1 - x^2) 
\bigg],
\label{sig:ell-gamma}
\end{equation}
where $x = \cos \theta$ and
$\sigma_T$ and $\sigma_L$ are the cross sections for a transverse 
and a longitudinal virtual photon, respectively.
The polar axis that defines the angle $\theta$ can be identified
with the SQA of the virtual photon.  
Our question can then be restated:  
for which SQA is the virtual photon 
most strongly polarized?

The longitudinal polarization 4-vector $\epsilon_L$ 
for a virtual photon with 4-momentum $Q$ must satisfy
$Q \cdot \epsilon_L = 0$ and $\epsilon_L^2 = -1$.
The most general 4-vector 
satisfying these conditions can be written in the form
\begin{equation}
\epsilon_L^\mu = \tilde X^\mu/\sqrt{-\tilde X^2}, \  \  
\tilde X^\mu = (-g^{\mu \nu} +  Q^\mu Q^\nu/Q^2) X_\nu,
\label{epsilonL-Xtilde}
\end{equation}
where  $X$ is a 4-vector.  The physical interpretation 
of $X$ is that in the dilepton rest frame,
which is defined by $\bm{Q} = 0$,  
the direction of the 3-vector $-\bm{X}$ is the SQA
of the virtual photon.

We now consider the production of a dilepton in the collision 
of two hadrons with 4-momenta $P_1$ and $P_2$. 
The SQA is generally chosen 
to lie in the production plane defined by the momenta 
of the colliding hadrons in the dilepton rest frame. 
Thus $X$ in Eq.~(\ref{epsilonL-Xtilde}) has the form
\begin{equation}
X^\mu = a P_1^\mu + b P_2^\mu,
\label{X-P1P2}
\end{equation}
where $a$ and $b$ are scalar functions. 
In the dilepton rest frame, the SQA is
antiparallel to the unit vector
$\bm{\hat X} = 
(a \bm{P}_1 + b \bm{P}_2)/|a \bm{P}_1 + b \bm{P}_2|$,
so it is determined by the ratio $a/b$.
If the leptons are the only particles in the final state 
whose momenta are measured, then $a/b$ can  
only depend on the 4-vectors $P_1$, $P_2$, and $Q$.
If additional information about the final state is measured, 
$a/b$ can also depend on this information.  

If the invariant mass $Q \equiv (Q^2)^{1/2}$ of the dilepton is large 
compared to the scale $\Lambda_{\rm QCD}$ of nonperturbative effects 
in QCD, the inclusive cross section can 
be calculated using QCD factorization formulas. 
Suppose the cross section is dominated by a specific parton 
process which, at leading order in the QCD coupling constant, 
predicts that the virtual photon is transversely polarized 
for any choice of the SQA.  Then a natural 
prescription for an optimal SQA 
is that it minimizes the longitudinal cross section 
from that parton subprocess at leading order.  
Such an optimal SQA will depend on the longitudinal 
momentum fractions of the colliding partons, 
which cannot be directly measured.

We first consider the production of a dilepton with large 
invariant mass $Q \gg \Lambda_{\rm QCD}$.
The Drell-Yan mechanism is the parton process $q \bar q \to \gamma^*$.
If the momenta of the quark and antiquark
are collinear to those of their parent hadrons, the 
virtual photon is transversely polarized
for any choice of the SQA. 
A rigorous QCD calculation of the transverse momentum distribution 
for the dilepton requires the resummation of the effects of the emission 
of soft gluons from the colliding partons \cite{Collins:1984kg}.
One of the most important qualitative effects of soft-gluon emission 
is that it gives transverse momenta to the colliding partons.
A simple model for these effects is the parton model 
with intrinsic transverse momentum, in which 
the momentum distribution of a parton is the product
of the parton distribution and a function of the 2-dimensional 
transverse momentum vector $\bm{k}_\perp$. 
Collins and Soper calculated the angular distribution of the dilepton 
in this model \cite{Collins:1977iv}.
The cross section for $q \bar q \to \gamma_L^*$,
where $\gamma_L^*$ is a longitudinal virtual photon, is 
\begin{equation}
\hat \sigma_L = 
\frac{8 \pi^2 e_q^2 \alpha \langle k_\perp^2 \rangle 
	(a^2 x_2^2 + b^2 x_1^2)}{3 Q^2 (a x_2 - b x_1)^2}
\delta(x_1 x_2 s - Q^2),
\label{sigpart-ave}
\end{equation}
where $x_1$ and $x_2$ are the longitudinal momentum fractions
of the colliding partons and $e_q$ is the electric charge of the quark.
We have used the expression for the longitudinal
polarization vector in Eqs.~(\ref{epsilonL-Xtilde}) and (\ref{X-P1P2}).
The cross section has been expanded to second order in 
$\bm{k}_{1\perp}$ and $\bm{k}_{2\perp}$ 
and averaged over them using
$\langle k_{n\perp}^i \big \rangle = 0$ and
$\langle k_{n\perp}^i k_{n\perp}^j \rangle
	= \frac12 \langle k_\perp^2 \rangle \delta_\perp^{i j}$,
where $\delta_\perp^{i j}$ is the unit tensor in the transverse
dimensions. 

Collins and Soper proposed a convenient set of angles 
$\theta$ and $\phi$ for the lepton momentum in the 
dilepton rest frame \cite{Collins:1977iv}.  
Lam and Tung pointed out that the Collins-Soper axis is an optimal 
SQA for the Drell-Yan mechanism  \cite{Lam:1978pu}.
It minimizes the cross section for longitudinal virtual photons
in Eq.~(\ref{sigpart-ave}), 
thus making the leading-order prediction
that the virtual photon will be transversely polarized 
as robust as possible with respect to radiative corrections 
that generate transverse momenta for the colliding partons.
Minimizing the cross section in Eq.~(\ref{sigpart-ave})
with respect to the ratio $a/b$, we find
\begin{equation}
a/b\big|_{\rm CS} = - x_1/x_2 .
\label{rat-min:x}
\end{equation}
Under the assumption that the cross section
is dominated by the Drell-Yan mechanism $q \bar q \to \gamma^*$,
we can derive an expression for $x_1/x_2$ in terms of 
variables that can be directly measured.
At leading order in $\bm{k}_{n\perp}$,
the energy-momentum conservation condition is
$Q = x_1 P_1 + x_2 P_2$, which implies 
\begin{equation}
x_1/x_2\big|_{\rm CS} = 
Q \cdot P_2/Q \cdot P_1 .
\label{rat-min:PQ}
\end{equation}
Equations~(\ref{rat-min:x}) and (\ref{rat-min:PQ}) define
the Collins-Soper axis.  The corresponding 4-vector $X$ 
in Eq.~(\ref{X-P1P2}) becomes
\begin{equation}
X_{\rm CS}^\mu =  
\frac{P_1^\mu}{Q \cdot P_1} - \frac{P_2^\mu}{Q \cdot P_2} .
\label{epsilonL-CS}
\end{equation}
In the dilepton rest frame, the SQA is along 
$\bm{v}_1 - \bm{v}_2$, where the $\bm{v}_i$'s are the velocities 
of the colliding hadrons.

We now consider the production of a dilepton with 
large transverse momentum $Q_T \gg \Lambda_{\rm QCD}$.
At leading order in $\alpha_s$, 
the parton processes that create a virtual photon at large $Q_T$ are
$q \bar q \to \gamma^* g$, $q g \to \gamma^* q$, 
and $\bar q g \to \gamma^* \bar q$.  
For $Q_T \gg Q$, the cross section for a longitudinal virtual photon 
is suppressed by a factor of $Q^2/Q_T^2$.
The coefficient of $Q^2/Q_T^2$ depends on the choice of  SQA. 
Our prescription for the optimal SQA is to 
minimize the longitudinal cross section,
which makes the transverse polarization for fixed $Q^2$ 
increase as rapidly as possible with increasing $Q_T$. 
We will refer to the optimal SQA for the parton process 
$i j \to \gamma^* k$ as the {\it optimal $ij$ axis}.

We first determine the optimal $q \bar q$ axis. 
At leading order in $\alpha_s$, 
the dependence of the differential cross section for
$q \bar q \to \gamma_L^* g$ on $a$ and $b$ is 
\begin{eqnarray}
Q^0 \frac{d \hat \sigma_L }{d^3Q} &\propto& 
\frac{a^2 x_2^2 + b^2 x_1^2}
    {(a w_1 + b w_2)^2 - a b Q^2 s},
\label{sig:qqbar2}
\end{eqnarray}
where $w_1 = Q \cdot P_1$, $w_2 = Q \cdot P_2$, and $s$ 
is the center-of-momentum (c.m.) energy of the colliding hadrons.
There is a delta function constraint on these variables:
$2(x_1 w_1 + x_2 w_2) = x_1 x_2 s + Q^2$.
Minimizing Eq.~(\ref{sig:qqbar2}) with respect to $a/b$, we get
\begin{eqnarray}
\frac{a}{b}\Big|_{q \bar q \to \gamma^* g} &=& 
\frac{x_1^2 w_1^2 - x_2^2 w_2^2 + Z^{1/2}}
    {x_2^2 (2 w_1 w_2 - Q^2 s)} ,
\label{ab:qqbar}
\\
&& \hspace{-2cm}
Z = (x_1^2 w_1^2 + x_2^2 w_2^2)^2
                 - x_1^2 x_2^2 Q^2 s (4 w_1 w_2 - Q^2 s) .
\nonumber
\end{eqnarray}
This also gives the optimal $\bar q q$ axis.

We next determine the optimal $q g$ axis.
At leading order in $\alpha_s$, 
the dependence of the differential cross section 
for $q g \to \gamma_L^* q$ on $a$ and $b$ is 
\begin{eqnarray}
Q^0 \frac{d \hat \sigma_L}{d^3Q} &\propto&
\frac{ (a x_2 - b x_1)^2  + b^2 x_1^2}
    {(a w_1 + b w_2)^2 - a b Q^2 s}.
\label{sig:qg2}
\end{eqnarray}
Minimizing with respect to $a/b$, we get
\begin{eqnarray}
\frac{a}{b}\Big|_{q g \to \gamma^* q} &=& 
\frac{2 x_1^2 w_1^2 - x_2^2 w_2^2
	+ Z^{1/2}}
    {x_2 (2 x_1 w_1^2 + 2 x_2 w_1 w_2 - x_2 Q^2 s)},
\label{ab:qg}
\\
&& \hspace{-2cm}
Z = (2 x_1^2 w_1^2 + x_2^2 w_2^2 
	           + 2 x_1 x_2 w_1 w_2 - x_1 x_2 Q^2 s)^2
\nonumber \\
&& 
               - x_1^2 x_2^2 Q^2 s (4 w_1 w_2 - Q^2 s).
\nonumber
\end{eqnarray}
This also gives the optimal $\bar q g$ axis.
The values of $a/b$ for the optimal $g  q$ and $g \bar q$ axes 
are obtained by taking the reciprocal of the expression 
on the right side of Eq.~(\ref{ab:qg}) and then interchanging 
$x_1$ and $w_1$ with $x_2$ and $w_2$.

The expressions for $a/b$ in Eqs.~(\ref{ab:qqbar}) and (\ref{ab:qg})
depend on the longitudinal momentum fractions $x_1$ and $x_2$ 
of the colliding partons only through the ratio $x_1/x_2$.
Our optimality criterion was based 
on the assumption that a specific parton process dominates.  
If that parton process implies a value for $x_1/x_2$
that depends on a measurable property of the final state, 
we can insert that value into Eqs.~(\ref{ab:qqbar}) and (\ref{ab:qg})
to obtain optimal SQA's that are experimentally useful. 
In the leading-order parton processes for a dilepton 
with large $Q_T$, the large transverse momentum is
balanced by that of a single parton.  This recoiling parton 
produces a jet of hadrons whose momenta are nearly collinear 
to that of the parton.  The polar angle of the recoiling parton 
in the hadron c.m.\  frame 
is approximately equal to the polar angle $\theta_{\rm jet}$ of the jet. 
The ratio $x_1/x_2$ can be expressed as a function of 
$\theta_{\rm jet}$ and the
transverse and longitudinal momenta $Q_T$ and $Q_L$ 
of the dilepton in the hadron c.m.\ frame: 
\begin{equation}
\frac{x_1}{x_2}\Big|_{\rm optimal} = 
\frac{(E_{\gamma^*} + Q_L) \sin \theta_{\rm jet}
	+ Q_T (1 + \cos \theta_{\rm jet})}
    {(E_{\gamma^*} - Q_L) \sin \theta_{\rm jet}
	+ Q_T (1 - \cos \theta_{\rm jet})} ,
\label{ratio-opt}
\end{equation}
where $E_{\gamma^*}= (Q_T^2 + Q_L^2 + Q^2)^{1/2}$.
Inserting this into Eqs.~(\ref{ab:qqbar}) and (\ref{ab:qg}),
we obtain expressions for $a/b$ for the optimal $q \bar q$ and $q g$ axes 
that depend on quantities that can be directly measured.

Beyond leading order in $\alpha_s$, there can be more than one 
parton in the final state, so there can be more than one jet 
with large transverse momentum.  As a general prescription for 
$x_1/x_2$, we choose $\theta_{\rm jet}$ in Eq.~(\ref{ratio-opt})
to be the angle in the hadron c.m.\ frame of the jet 
with the largest transverse energy.
Since the direction of the jet
is insensitive to soft gluon radiation and to the 
splitting of a parton into collinear partons,
QCD radiative corrections to the polar angle distributions 
defined by our optimal SQA's can be calculated 
systematically using perturbative QCD. 
The number of jets in an event depends on the size of the angular cone 
used to define the jet.  If a reasonably large value is chosen 
for the cone size, most of the events will contain a single jet 
with large transverse momentum balancing that of the dilepton.
A small fraction of the events will have more than one such jet. 
For these multijet events,
the expression on the right side of Eq.~(\ref{ratio-opt})
may not be a good estimate for $x_1/x_2$.
However since these events are only a small fraction 
of the total number of events, they should not seriously 
decrease the transverse polarization.
Our expression for $x_1/x_2$ is not useful for
fixed-target experiments, because the recoiling jet 
is usually not observed.  However  $\theta_{\rm jet}$
can be measured relatively easily in high energy hadron colliders.

To illustrate the use of our optimal SQA's,
we apply them to the production of a dilepton 
with invariant mass $Q=3$~GeV at the Tevatron,
which is a $p \bar p$ collider with c.m.\ 
energy 1.96~TeV.
We calculate the cross sections for a longitudinal virtual photon 
using the CTEQ6L parton distributions \cite{Pumplin:2002vw} 
with 3 flavors ($u$, $d$, and $s$).
The factorization and renormalization scales
are set to $\mu=(Q^2+Q_T^2)^{1/2}$.
For the strong coupling constant $\alpha_s(\mu)$,
we use the next-to-leading order formula 
with 4 flavors of quarks and
$\Lambda_\textrm{QCD}$=326~MeV.
For the electromagnetic coupling constant $\alpha$, we use
$1/132.6$.
We impose a rapidity cut $|y|<1$ on the dilepton momentum.

We compare our optimal SQA's with three other 
choices for the SQA:\\
\renewcommand{\theenumi}{\roman{enumi}}
\renewcommand{\labelenumi}{(\theenumi)}
\begin{enumerate}
\vspace{-4.5ex}
\item the {\it c.m.\ helicity axis} defined by $X_{\rm cmh} = P_1 + P_2$.
The projection of the spin of the virtual photon along this 
axis is minus its helicity in the hadron c.m.\ frame.\\ 
\vspace{-4.5ex}
\item
the {\it Collins-Soper axis} defined by Eq.~(\ref{epsilonL-CS}). \\  
\vspace{-4.5ex}
\item
the {\it perpendicular helicity axis} defined by 
\end{enumerate}
\vspace{-2ex}
\begin{equation}
X_{\perp {\rm h}}^\mu =  
\frac{P_1^\mu}{Q \cdot P_1} + \frac{P_2^\mu}{Q \cdot P_2} .
\label{epsilonL-perp}
\end{equation}
In the dilepton rest frame, the perpendicuclar helicity axis is along 
$\bm{v}_1 + \bm{v}_2$.
The spin of the virtual photon along this axis
coincides with its helicity in the frame obtained from the
hadron c.m.\ by a longitudinal boost that makes the dilepton
momentum perpendicular to the beam direction.
The perpendicular helicity axis coincides with the
c.m.\ helicity axis when the rapidity of the dilepton is 0.
One advantage shared by $X_{\perp {\rm h}}$ and 
$X_{\rm CS}$ is that they are invariant 
under independent longitudinal boosts 
of the two colliding hadrons.  Thus they can be expressed 
equally well in terms of the 
colliding parton momenta or the colliding hadron momenta.

\begin{figure}[t]
\includegraphics*[height=8cm,angle=0,clip=true]{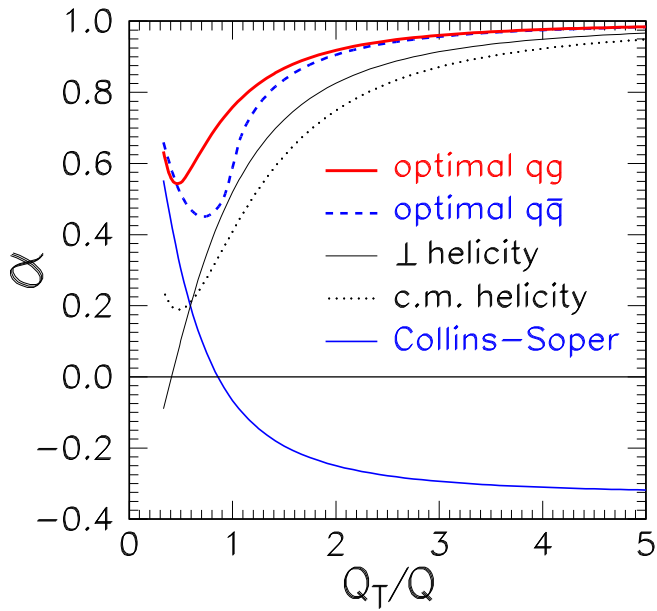}
\vspace*{-5ex}
\caption{(color online). 
Polarization variable $\alpha$ for various spin-quantization axes
as functions of $Q_T/Q$ for dileptons with invariant mass
$Q = 3$~GeV at the Tevatron.
\vspace*{-2ex}
}
\label{fig:alphaTev}
\end{figure}
%

A convenient polarization variable for a virtual photon is 
$\alpha = (\sigma_T - 2 \sigma_L)/(\sigma_T + 2 \sigma_L)$.
It can be measured from the polar angle distribution of the lepton, 
which is proportional to $ 1 + \alpha \cos^2 \theta$.
The leading-order predictions for $\alpha$
for various choices of SQA  
are shown as functions of $Q_T/Q$ in Fig.~\ref{fig:alphaTev}. 
For the Collins-Soper axis, $\alpha$ is negative for $Q_T > Q$. 
For the c.m.\ helicity axis, the perpendicular helicity axis, 
and our two optimal axes, $\alpha$ approaches 1 at large $Q_T$, 
indicating pure transverse polarization. 
The approach to 1 is significantly faster for the 
perpendicular helicity axis than for the c.m.\ helicity axis.  
However it is even 
faster for both of the optimal axes.  The two optimal axes 
give essentially the same $\alpha$ for $Q_T>2Q$, but for $Q_T < 2Q$
$\alpha$ is closer to 1 for the  
optimal $q g$ axis than for the optimal $q \bar q$ axis.

The requirement that the recoiling jet be captured by the 
hadronic calorimeter imposes a constraint on its 
pseudorapidity $\eta_{\rm jet} = \ln \tan(\theta_{\rm jet}/2)$.
To provide some idea of how much this additional constraint
might decrease the data sample, we impose the pseudorapidity cut 
$|\eta_{\rm jet}| < 1$.
For dileptons with $Q=3$~GeV at the Tevatron,
the fraction of events satisfying the dilepton rapidity cut 
that also survive the jet pseudorapidity cut is greater than
0.5 for $Q_T > Q$.
This fraction is large enough that measuring the angle 
$\theta_{\rm jet}$ 
should not dramatically decrease the size of the data sample. 

We have also compared the various quantization axes
for dilepton production at the LHC,
which is a $p p$ collider with c.m.\  energy 14~TeV.
We imposed a rapidity cut $|y|<3$ on the dilepton.
For $Q_T>Q$, the leading-order results for $\alpha$ differ by
less than 0.01 from the results for the Tevatron in
Fig.~\ref{fig:alphaTev} for all the SQA's with one exception.
For the c.m.\ helicity axis, $\alpha$ at the LHC
is smaller by more than 0.1 in the region near $Q_T = 1.5~Q$.
The leading-order predictions for $\alpha$ as functions of $Q_T/Q$ are
similar to those for the Tevatron in Fig.~\ref{fig:alphaTev}.
If we impose a pseudorapidity cut $|\eta_{\rm jet}| < 3$
on the recoiling jet,
the fraction of events satisfying the dilepton rapidity cut
that also survive the jet pseudorapidity cut is greater than
0.8 for $Q_T > Q$.

The CDF and D0 Collaborations have measured 
$\alpha$ as a function of $Q_T$ for dimuons produced at the 
Tevatron from decays of charmonium mesons 
\cite{CDF:polznpsi}
and bottomonium mesons \cite{Acosta:2001gv,D0:2008za}.
The variable $\alpha$ for the c.m.\ helicity axis
was measured for $Q_T/Q$ as high as 9.7 for $J/\psi$ 
and 2.1 for $\Upsilon(1S)$.
The data samples collected by CDF and D0 are large enough 
that it should also be possible to measure $\alpha$
for the sidebands of these quarkonium resonances,
which come from dileptons produced by virtual photons. 
At the LHC, it should be possible 
to measure $\alpha$ out to much larger values of $Q_T/Q$.

Our leading-order results for $\alpha$ 
as a function of $Q_T$ indicate that the virtual photon will be 
significantly more strongly polarized at large transverse momentum
if we use optimal SQA's.  A quantitative prediction of the polarization
requires calculations to next-to-leading order (NLO) in $\alpha_s$.  
The angular distributions for lepton pairs at large $Q_T$ 
have been calculated to NLO by Mirkes and Ohnemus \cite{Mirkes:1994dp}.
At NLO, there can be two partons in the final state.
In Ref.~\cite{Mirkes:1994dp}, the momenta of the two partons 
were integrated over, so they cannot be resolved into two 
separate jets.  In order to use these NLO results,
we would have to modify our prescription for $x_1/x_2$.
At small $Q_T$, it is also necessary to sum up the effects 
of soft-gluon emission to all orders \cite{Collins:1984kg}.

Similar methods could be used to derive optimal SQA's  
for the production at large transverse momentum of 
massive particles, such as $J/\psi$, $\Upsilon$,
and the weak bosons $W^\pm$ and $Z^0$.
With an optimal SQA, these states should be
more strongly polarized, so measurements of 
$\alpha$ will provide more information.
It may also make the theoretical predictions more robust 
with respect to radiative corrections.
Optimal SQA's might be useful for determining 
the spin of new particles created at the LHC.  
They provide a new window into the spins of particles
created by parton collisions.

\begin{acknowledgments}
This research was supported in part by the Department of Energy 
under grants DE-FG02-05ER15715 and DEFC02-07ER41457, 
by the KOSEF under grant R01-2008-000-10378-0,
and by the Korea Research Foundation under grant KRF-2006-311-C00020.
\end{acknowledgments}

{}

\end{document}